\documentclass[amsmath, amssymb, superscriptaddress, reprint, twocolumn, colorlinks=true, citecolor=blue, linkcolor=blue, urlcolor=blue]{revtex4-1}
\usepackage{natbib}
\usepackage[utf8]{inputenc}
\usepackage[T1]{fontenc}
\usepackage{graphicx}
\usepackage{grffile}
\usepackage{longtable}
\usepackage{wrapfig}
\usepackage{rotating}
\usepackage[normalem]{ulem}
\usepackage{amsmath}
\usepackage{textcomp}
\usepackage{amssymb}
\usepackage{capt-of}
\usepackage{hyperref}
\usepackage{mathptmx}
\usepackage{upgreek}
\usepackage{float}
\usepackage{bm}
\usepackage{booktabs}
\usepackage{subfigure}
\usepackage{url}
\usepackage{tablefootnote}

\begin{document}

\title{Abnormal Polarity Effect on the DC Breakdown Voltage in Short SF\(_6\) Gap}

\author{Zihao Feng}
\affiliation{Department of Electrical Engineering, Tsinghua University, Beijing 100084, China}

\begin{abstract}

In this Letter, through the comparison between experiment and numerical simulation, we reveal the dynamic mechanism underlying the abnormal polarity effect in SF\(_6\) short-gap DC breakdown, as well as a novel criterion for predicting negative breakdown voltage. Using the traditional single-streamer breakdown criterion, the simulated positive breakdown voltage agrees well with experimental measurements, whereas the simulated negative breakdown voltage deviates markedly from the experiments, so the single-streamer breakdown criterion fails to reproduce the abnormal polarity effect observed experimentally. In addressing this, we propose an ion–ion plasma breakdown criterion for negative breakdown voltage. When this novel criterion is applied, the simulated negative breakdown voltage agrees with the experiments and reflects the abnormal polarity effect. Analysis of the spatiotemporal evolution of key physical parameters reveals that, the dynamic mechanism for ion-ion plasma breakdown for negative polarity can be divided into four stages: primary streamer stage, ion accumulation stage, reconstructive ionization stage, and ion–ion plasma propagation stage. Notably, the ion–ion plasma propagation stage is dominated by photoionization-driven negative-ion accumulation rather than conventional impact ionization.

\end{abstract}

\maketitle

\begin{figure*}[t]
\centering
\includegraphics[width=17cm]{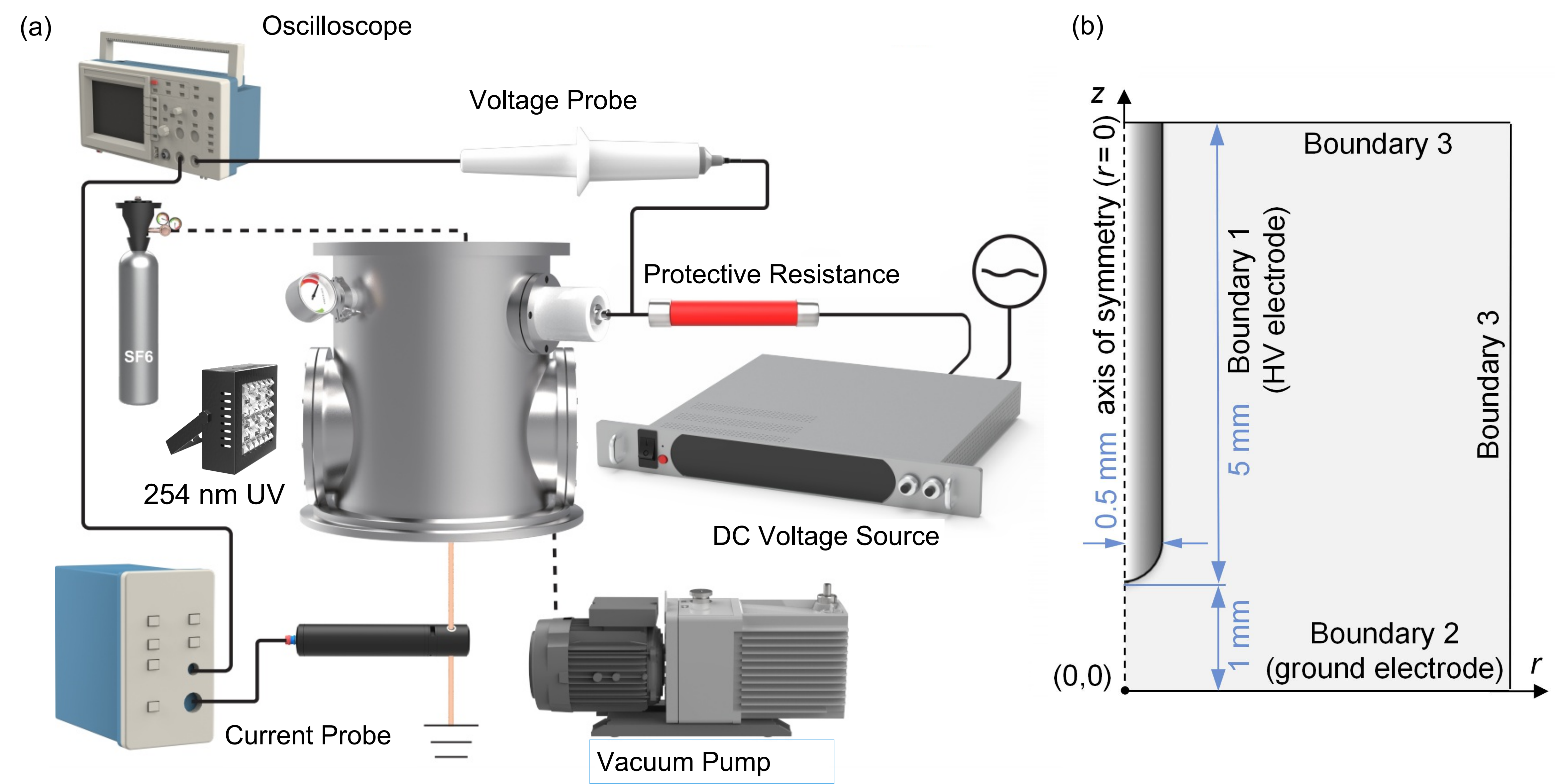}
\caption{\label{fig.1} (a) Experimental system for measuring the DC breakdown voltage. (b) Rod-plane discharge electrode and computational boundaries.}
\end{figure*}

DC breakdown is the most common form of electrical breakdown in gas-insulated electrical equipment\cite{https://doi.org/10.1002/tee.24244,10.1063/5.0223522,10007898,Francisco_2021}. In particular, discharges of different polarities may occur simultaneously in the context of metal-particle-induced discharges \cite{arxiv123,8928273,10.1088/1361-6463/add944}. Therefore, examining the polarity effect on DC breakdown voltage is crucial for advancing the quantitative analysis of these issues. The polarity effect on the DC breakdown voltage in short gaps refers to the phenomenon where breakdown voltages differ significantly for different DC polarities under highly non-uniform fields. The conventional polarity effect (i.e., \(U_\text{positive}<U_\text{negative}\)) has been well-explained in extensive textbooks \cite{FDLFSD,gaodianya,shao2015gas} from the perspective of positive-ion dynamics within the framework of streamer mechanism. The conventional polarity effect is well validated in air gap breakdown. However, for the strongly electronegative SF\(_6\) gas, the polarity effect is entirely different, i.e., the short-gap DC breakdown voltage for SF\(_6\) typically exhibits \(U_\text{positive}>U_\text{negative}\). This abnormal polarity effect has already been observed in some experimental measurements. For instance, Seeger \textit{et al.} (see Fig.2 in Ref.\cite{Seeger_2014}) employed a 254 nm UV lamp to reduce the statistical time lag and measured the so-called maximum DC breakdown voltage (max BD) for SF\(_6\) rod-plane gap. This max BD corresponds to breakdown caused solely by the streamer mechanism, without the leader mechanism. They observed that the abnormal polarity effect \(U_\text{positive}>U_\text{negative}\) occurs in SF\(_6\) at pressures \(P \geq 1 \ \text{atm}\). The traditional positive-ion dynamics cannot explain this abnormal polarity effect, and to date, the dynamic mechanism or quantitative verification for explaining the SF\(_6\) abnormal polarity effect remains limited, which motivates the research presented in this Letter. Notably, the above discussion focuses on breakdown in short gaps, solely involving the streamer mechanism. In contrast, long-gap breakdown involves both streamer and leader mechanisms \cite{Seeger_2009,https://doi.org/10.1029/2023GL102834,https://doi.org/10.1049/hve2.70076,Wu_2019}. However, since the streamer mechanism plays a crucial role in leader initiation \cite{Seeger_2008123,SF6111,https://doi.org/10.1049/hve2.12119,Zhao_2022}, it is necessary to decouple these two mechanisms and first investigate the abnormal polarity effect in the framework of streamer mechanism using short gaps, which is the research objective of this Letter.

In this Letter, we employ experimental measurements and 2D plasma fluid simulations to quantitatively compare the DC breakdown voltage of SF\(_6\) in short gaps. Initially, simulations utilizing a single-streamer breakdown criterion show that the negative breakdown voltage does not match the experimental results and fail to reflect the abnormal polarity effect. Subsequently, we propose a novel criterion for negative breakdown voltage, i.e., ion-ion plasma breakdown criterion, through which the simulated negative breakdown voltage shows quantitative agreement with experimental measurements and exhibits the abnormal polarity effect (\(U_\text{positive}>U_\text{negative}\)). Finally, on the basis of the simulation results, we explain in detail the dynamic processes involved in the ion-ion plasma breakdown criterion under negative DC voltage and thus, explain the abnormal polarity effect.

The experiment and simulation both use a rod-plane electrode to create a 1 mm short gap in a highly non-uniform field, with SF\(_6\) gas at 1 atm, as shown in Fig. \ref{fig.1}. The simulation employs the classical fluid model under the local field approximation. Notably, photoionization plays a non-negligible role in the quantitative prediction of SF\(_6\) breakdown voltage. The photoionization source terms in the continuity equations are calculated using the computational photoionization model and three-group Helmholtz parameters for 1 atm SF\(_6\) proposed by Feng \textit{et al.} \cite{arxiv}, in which the theoretical foundation is based on Zheleznyak's classical photoionization model \cite{ZK} and Pancheshnyi's analytical model \cite{Pancheshnyi_2015}, while the numerical computation follows the Helmholtz equation model proposed by Luque \textit{et al.} \cite{10.1063/1.2435934} (and in parallel by Bourdon \textit{et al.} \cite{Bourdon_2007}). The detailed description of the experimental setup and numerical model are provided in the Supplementary Material of this Letter.

\begin{figure}[t]
\centering
\includegraphics[width=8.5cm]{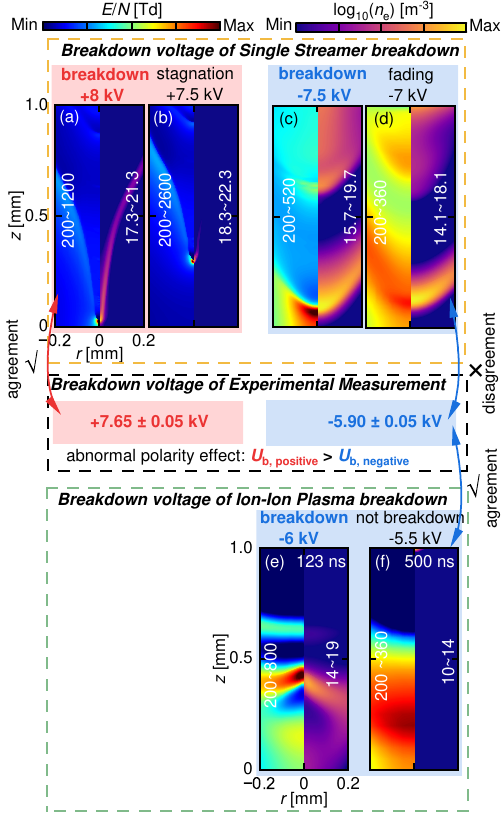}
\caption{\label{fig.2} Comparison between experimental and simulated DC breakdown voltages. (a–d) Simulated breakdown voltages predicted using the single-streamer breakdown criterion. (e–f) Simulated breakdown voltages predicted using the ion–ion plasma breakdown criterion. The experimentally measured breakdown voltages are also indicated in the figure.}
\end{figure}

The experimental results, presented in Fig. \ref{fig.2}, show a negative breakdown voltage of -5.9 kV and a positive breakdown voltage of 7.65 kV. This clearly demonstrates a abnormal polarity effect, where \(U_\text{positive}>U_\text{negative}\), which qualitatively consistent with previous experimental results reported by Seeger \textit{et al.} (see Fig.2 in Ref.\cite{Seeger_2014}). 

In the simulation, initially, we apply the single-streamer breakdown criterion. For positive breakdown, according to Refs. \cite{BKDEA,Niknezhad_2021,Li_2022}, if breakdown does not occur, the positive streamer stagnates, as shown in Fig. \ref{fig.2}(b). The electric field rapidly increases from 1500 Td to 3000 Td within 0.1 ns, exceeding the drift-diffusion approximation's limits. The electron density reaches \(10^{23} \, \text{m}^{-3}\), which is too high for a non-equilibrium, weakly ionized plasma. This non-physical behavior arises from the local source term expression used in the classical fluid model under the local field approximation, indicating that the positive streamer stops developing under these conditions. Therefore, the positive breakdown voltage is identified as 8 kV. For negative breakdown, according to Refs. \cite{BKDEA,Guo_2022123}, if breakdown does not occur, the negative streamer fades, and the electric field at the streamer head drops below the effective ionization threshold \((E/N)_\text{cr} = 360 \ \text{Td}\) for SF\(_6\), identifying the negative breakdown voltage as -7.5 kV.

Comparing experimental results with the simulated breakdown voltages predicted by the single-streamer criterion shows agreement for positive polarity, validating the criterion. However, for negative polarity, the simulated breakdown voltage is significantly higher than the experimental value, highlighting a discrepancy. Thus, the single-streamer breakdown criterion fails to capture the abnormal polarity effect in SF\(_6\), suggesting that additional mechanisms may be involved in negative polarity breakdown, leading to a lower negative breakdown voltage.

The following sections introduce a novel criterion for SF\(_6\) negative breakdown. After the negative streamer enters the fading state, we continue the simulation, revealing a series of surprising phenomena that explain the abnormal polarity effect in SF\(_6\). This is termed the "ion-ion plasma breakdown criterion" in this Letter. 

First, consider the breakdown voltage predicted by ion-ion plasma breakdown criterion. In the simulation with an applied voltage of -6 kV, the ion-ion plasma sustains ionization from the rod electrode to the plane electrode, bridging the gap, as shown in Fig. \ref{fig.2}(e). Although the electron density is only  \(10^{18} \, \text{m}^{-3}\), much lower than typical streamer discharge, and the electric field of the ionization front is below the \((E/N)_\text{cr} = 360 \ \text{Td}\), the breakdown still occurs under this scenario. This suggests a potential ionization mechanism distinct from the impact ionization, which will be detailed explained later. In the simulation with an applied voltage of -5.5 kV, as shown in Fig. \ref{fig.2}(f), no ionization structure bridges the gap, and the electron density drops below \(10^{16} \, \text{m}^{-3}\), resulting in insufficient conductivity, thus breakdown does not occur. This behavior resembles negative corona discharge \cite{15}, with corona drifting to the grounded electrode at around 500 ns, rather than ionization. Thus, -6 kV is identified as the negative breakdown voltage using the ion-ion plasma breakdown criterion, matching experimental measurements in Fig. \ref{fig.2}. It also reflects the abnormal polarity effect, with \(U_\text{positive}>U_\text{negative}\).

\begin{figure*}[t]
\centering
\includegraphics[width=17cm]{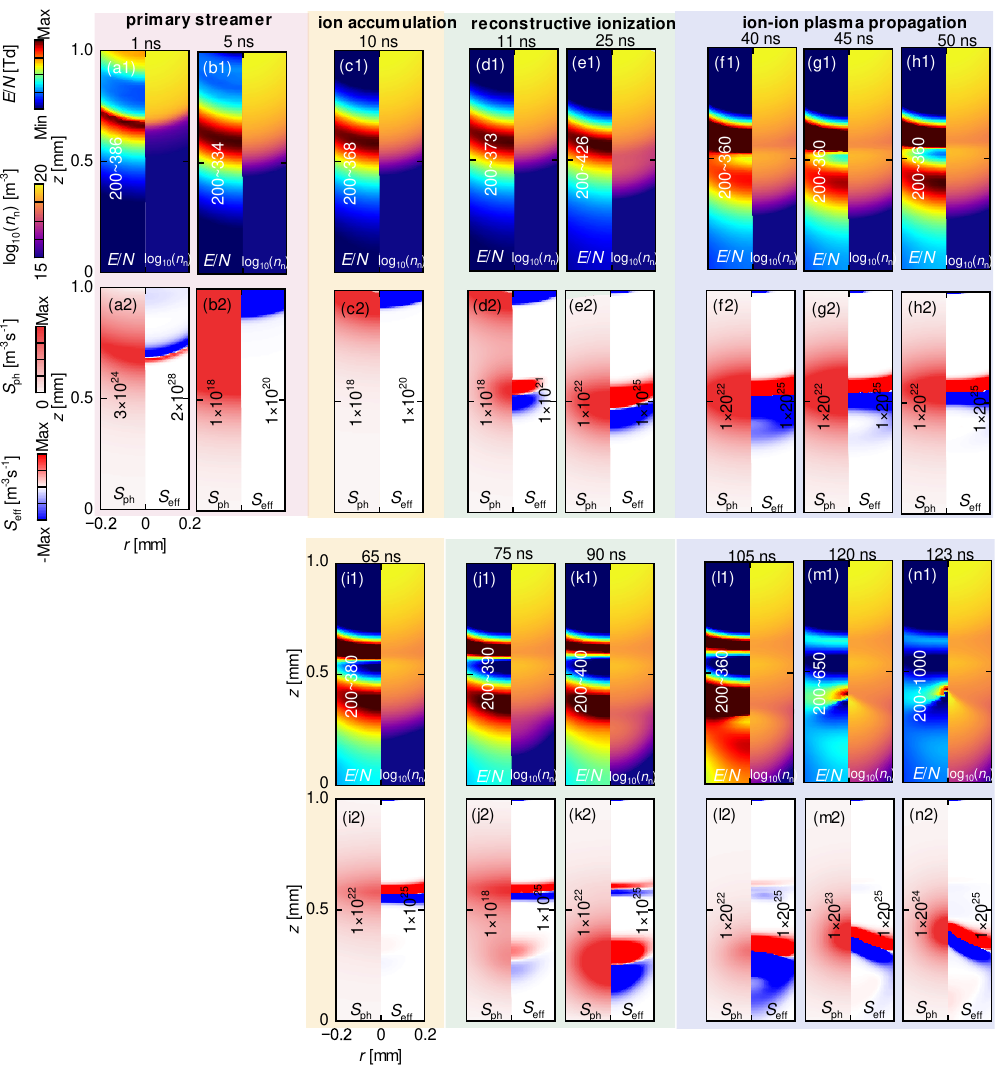}
\caption{\label{fig.3} Spatiotemporal evolution of the reduced electric filed \(E/N\), the logarithm of negative ion density \(\log_{10}(n_n)\), photoionization rate \(S_\text{ph}\) and effective ionization rate \(S_\text{eff}\), illustrating the dynamic mechanism of ion-ion plasma breakdown. Accordingly, the process is divided into four stages: (a–b) the primary streamer stage, (c) and (i) the ion accumulation stage, (d–e) and (j–k) the reconstructive ionization stage, and (f–h) and (l–n) the ion–ion plasma propagation stage.}
\end{figure*}

For explaining the dynamic mechanism of ion-ion plasma breakdown, we analyze the spatiotemporal evolution of key physical parameters for the -6 kV case in Fig. \ref{fig.3}, including the reduced electric field \(E/N\), the logarithm of negative ion density \(\log_{10}(n_n)\), photoionization rate \(S_\text{ph}\), and effective ionization rate \(S_\text{eff}\). Note that \(S_\text{ph}\) only relates to impact ionization, while \(S_\text{ph}\) is influenced by both impact ionization and attachment. Based on the dynamic evolution, we divide the ion-ion plasma breakdown into four stages: the primary streamer stage, ion accumulation stage, reconstructive ionization stage, and ion-ion plasma propagation stage.

During the primary streamer stage, as shown in Fig. \ref{fig.3}(a-b), the negative streamer remains propagation at 1 ns. By 5 ns, it enters the negative streamer fading state. At this point, both the \(S_\text{eff}\) and \(S_\text{ph}\) significantly decrease, halting ionization and transforming the negative streamer into an ion-ion plasma.

Subsequently, the discharge enters the ion accumulation stage, as shown in Fig. \ref{fig.3}(c). Specifically, under the negative applied voltage, electrons generated during the past primary streamer stage drift forward and undergo attachment, forming a region dominated by negative ions at the front of the ion-ion plasma. The space charge field of this negative ion region aligns with the negative applied field, causing the front field of the ion-ion plasma exceeding the \((E/N)_\text{cr} = 360 \ \text{Td}\) and reaching 368 Td at 10 ns. But \(S_\text{ph}\) and \(S_\text{eff}\) still remain rather weak.

At 11 ns, at the front of the ion-ion plasma, \(S_\text{ph}\) and \(S_\text{eff}\) recover to a non-negligible level, marking the transition to the reconstructive ionization stage, as shown in Fig. \ref{fig.3}(d-e). It is important to note that, although the front electric field exceeds \((E/N)_\text{cr}\) at this stage, new ionization structure does not occur immediately. This is due to the significant decrease in electron density \(n_\text{e}\) during the past ion accumulation stage; for example, at 11 ns, the \(n_\text{e} <10^{16} \, \text{m}^{-3}\). According to the expressions for \( S_\text{ph} \) and \( S_\text{eff} \), \( S_\text{ph} \) is positively correlated with \( \alpha n_{\mathrm{e}} \mu_{\mathrm{e}} E \), while \( S_{\text{eff}}=(\alpha - \eta) n_{\mathrm{e}} \mu_{\mathrm{e}} E\). Thus under conditions of low \(n_\text{e}\), even with a high electric field, the \( S_\text{ph} \) and \( S_\text{eff} \) remain insufficient to reach a significant high level. Therefore, before new ionization structure can be reconstructed, the discharge undergoes a finite evolution, during which residual electrons (\(<10^{16} \, \text{m}^{-3}\)) undergo electron avalanche, until it increases to a high level (\(>10^{16} \, \text{m}^{-3}\)) that can support high \( S_\text{ph} \) and \( S_\text{eff} \), ultimately forming a pronounced new ionization structure shown in Fig. \ref{fig.3}(e). Notably, we would remark that, the above-mentioned process is fundamentally different from the so-called bipolar ion drift, which is argued to be the dominated factor for precursor mechanism in stepped discharge propagation (see Section. 2.3.2 in Ref. \cite{Seeger_2009}). The reason is, if bipolar ion drift were the trigger, new ionization structure would appear within the negative ion region, where both ion mobility and ion density are highest. However, as shown in Fig. \ref{fig.3}(e1), new ionization structure forms ahead of this negative ion region, rather than within it.

Finally, the discharge enters the ion-ion plasma propagation stage, which is dominated by photoionization-driven negative ion accumulation, as shown in Fig. \ref{fig.3}(f-h). Notably, the ion-ion plasma at this stage comprises a positive streamer propagating backwards towards the negative high-voltage electrode (a reverse ionization wave) and a negative-ion region, expanding forward towards the grounded electrode. The formation of the positive streamer in this Letter (see Fig. \ref{fig.3}(f-h) and Fig. \ref{fig.3}(l-n)) is similar to the phenomenon reported in Ref. \cite{nature}, where a wide screening-ionization wave collapses, leading to a sudden transition into the positive streamer. In the ion-ion plasma propagation stage, the reverse positive streamer serves to supplement the electrons in the channel, enhancing conductivity, but this role is secondary for the gap breakdown. The primary role for the gap breakdown is the forward-expanding negative-ion region. Notably, the electric field within this forward-expanding negative-ion region remains below \((E/N)_\text{cr}\). Consequently, the forward-expanding region is not dominated by impact ionization but rather by photoionization. Specifically, the photoionization rate \( S_\text{ph} \) is positively correlated with \( \alpha n_{\mathrm{e}} \mu_{\mathrm{e}} E \), where \( \alpha \) denotes the impact ionization coefficient. Thus photoionization rate depends solely on the impact ionization and is independent of the attachment. Therefore, even when the electric field is below \((E/N)_\text{cr}\), photoionization remains active. Furthermore, photoionization is inherently non-local, thus it continuously generates seed photoelectrons at the front edge of the negative-ion region. But \( S_{\text{eff}}=(\alpha - \eta) n_{\mathrm{e}} \mu_{\mathrm{e}} E\), where \( \eta \) denotes the attachment coefficient. At the front edge of the negative-ion region, the electric field is below \((E/N)_\text{cr}\), so \(S{_\text{eff}} < 0 \), meaning attachment surpasses impact ionization. As a result, photoelectrons are rapidly attached to form negative ions at the front edge of the negative-ion region, causing the ion-ion plasma propagation. 

The underlying physical mechanism of the ion-ion plasma propagation is, in essence, consistent with the so-called "passive propagation mode of the primary streamer" reported in Ref. \cite{arxiv123} for high-pressure SF\(_6\). However, a difference lies in the fact that under the high-pressure conditions of Ref. \cite{arxiv123}, so-called SF\(_6\) side streamers will occur, and the forward-expanding negative-ion region can be enhanced by merging with these side streamers, thereby increasing the negative space charge density. Consequently, one can conclude that the ion-ion plasma breakdown criterion proposed in this Letter also validates at higher pressures (\(P > 1 \ \text{atm}\)) and will be enhanced by the presence of SF\(_6\) side streamers, potentially leading to an even more pronounced reduction in the negative breakdown voltage.

The ion accumulation stage, reconstructive ionization stage, and ion-ion plasma propagation stage can repeat during a single discharge event, continuing until ion-ion plasma bridges the anode and cathode, leading to gap breakdown, as shown in Fig. \ref{fig.3}(i-n).

In conclusion, through the experimental and numerical investigation, this Letter provides a novel criterion for the SF\(_6\) negative DC breakdown voltage prediction and reveals the dynamic mechanism for the abnormal polarity effect in SF\(_6\) short-gap DC breakdown. The experimental results show that the negative breakdown voltage is significantly lower than positive breakdown voltage, exhibiting the abnormal polarity effect (i.e., \( U_{\text{positive}} > U_{\text{negative}} \)). In the simulation, the breakdown voltage is initially predicted using the single-streamer breakdown criterion. For positive polarity, the single-streamer breakdown voltage agrees with the experimental measurements. However, for negative polarity, the single-streamer breakdown voltage is significantly higher than the experimental value, and even exceeds the positive breakdown voltage. This indicates that the single-streamer criterion is insufficient to reflect the abnormal polarity effect. Therefore, we propose a novel breakdown criterion, i.e., the ion-ion plasma breakdown criterion, for SF\(_6\) negative breakdown voltage. The negative breakdown voltage predicted using this novel criterion is consistent with the experimental measurements and reflects the abnormal polarity effect in SF\(_6\) short gap. A detailed analysis of the plasma dynamics for SF\(_6\) negative polarity breakdown reveals that the ion-ion plasma breakdown consists of four stages: primary streamer stage, ion accumulation stage, reconstructive ionization stage, and ion-ion plasma propagation stage. The comprehensive dynamic processes of each stage, along with the dominated physical parameters, have been identified. Notably, the ion–ion plasma propagation stage is dominated by photoionization-driven negative-ion accumulation rather than conventional impact ionization.

\section*{Acknowledgment}
The authors gratefully acknowledge funding support from the National Natural Science Foundation of China (Grant No. 52407176) and the Postdoctoral Fellowship Program of CPSF (Grant No. GZB20230326).

\section*{AUTHOR DECLARATIONS}
Conflict of Interest. The authors have no conflicts to disclose.

\section*{SUPPORTING INFORMATION }
See the Supplementary Material for the detailed description of the experimental setup and numerical model.

\section*{DATA AVAILABILITY}
All data that support the findings of this study are included within the article (and any supplementary files).

\Large
\bibliography{references}

\end{document}